\documentclass{article}

\usepackage{graphicx}
\usepackage{psfig}
\usepackage{epsfig}
\usepackage[round]{natbib}

\title{
Statistical modelling of tropical cyclone tracks: modelling the
autocorrelation in track shape}

\begin{document}

\author{Tim Hall, GISS\footnote{\emph{Correspondence address}: Email: \texttt{tmh1@columbia.edu}}\\and\\
Stephen Jewson\\}

\maketitle

\begin{abstract}
We describe results from the third stage of a project to build a
statistical model for hurricane tracks. In the first stage we
modelled the unconditional mean track. In the second stage we
modelled the unconditional variance of fluctuations around the
mean. Now we address the question of how to model the
autocorrelations in the standardised fluctuations. We perform a
thorough diagnostic analysis of these fluctuations, and fit a type
of AR(1) model. We then assess the goodness of fit of this model in a
number of ways, including an out-of-sample comparison with a simpler model,
an in-sample residual analysis, and a comparison of simulated
tracks from the model with the observed tracks. Broadly speaking, the model captures
the behaviour of observed hurricane tracks. In detail, however, there are
a number of systematic errors.
\end{abstract}

\section{Introduction}

There are six distinct regions in which tropical cyclones occur: the
North Atlantic, the North East Pacific, the North West Pacific,
the South West Pacific, the Northern Indian Ocean and the Southern
Indian Ocean. Tropical cyclones in the North Atlantic (the
stronger ones of which are often known as hurricanes) affect the
Caribbean, Central America and the US. Those in the North East
Pacific (also known as hurricanes) can affect Mexico. Storms in
the North West Pacific (also known as typhoons) frequently impact
Japan, Taiwan, Korea and China. South Pacific storms may hit the east
coast of Australia or New Zealand. Northern Indian Ocean storms,
although infrequent, can make landfall in India or Bangladesh,
while Southern Indian Ocean storms can hit the West Coast of
Australia, and Indian ocean islands such as Madagascar.

Tropical cyclones in all of these regions have the potential to
cause significant damage and loss of life. As a result they affect many groups
of people in each of these communities including home-owners, businesses,
health-care providers and, in the wealthier countries, the
insurance industry. Many of these groups might benefit from a more
accurate assessment of the tropical cyclone risk they face, since a
more accurate understanding of the risk could lead to better
decisions about preventative measures. Making approximate risk
assessments is rather easy, since information about historical
tropical cyclones is widely available. However, many factors
conspire to make more accurate risk
assessments difficult. These include the fact that
the most severe tropical cyclones are rare,
that the geometry of islands and coastlines is often complicated,
which can make the comparison of wind records from nearby
locations difficult, and that long term
fluctuations in the climate, including anthropogenic climate
change, may be changing underlying activity rates (and these changes may be
different in different regions).

One class of methods for estimating tropical cyclone risk that reduces these difficulties
is the basin-wide track modelling approach, as followed
by~\citet{fujii}, \citet{drayton00}, \citet{vickery00} and~\citet{emanuel05}.
In this approach statistical models are constructed that
attempt to represent the behaviour of the tracks and intensities of
tropical cyclones across a whole basin. The potential advantages
of this approach are that it allows the use of all the available
track information for a particular basin, and that the total
impact of a storm can be calculated rather easily. The potential
disadvantage is that the modelling may be rather difficult
because of the complicated nature of tropical cyclone tracks. A poor
model may then give less accurate estimates of risk than a much
simpler approach that did not use basin-wide track modelling.

We are currently building a new basin-wide tropical cyclone model from scratch.
Because of the complex nature of the problem we are taking a systematic approach that
includes extensive validation and cross-validation of our models.
Our initial focus is on Atlantic hurricanes, and
so far we have completed two stages in the modelling process.
The first stage was to model the mean tracks of Atlantic hurricanes
(see~\citet{hallj05a}).
We do this by using a weighted nearest neighbour approach, in which the definition of `near' is optimised
using a jack-knife out-of-sample fitting procedure.
The second stage was to model the variance of fluctuations around these mean tracks
(see~\citet{hallj05b}), and again we use a weighted nearest neighbour approach, with optimal definition
of near. Putting together the first two stages, one can generate random artificial hurricane tracks,
and we show such tracks in figure 6 of~\citet{hallj05b}.
These tracks are clearly not realistic, however
(they are too jagged, or irregular, relative to real tracks), and it seems that the
most important missing element is along-track memory.
The third stage of our research, which we describe in this article,
is therefore to attempt to model this memory (or `inertia')
in the standardised anomalies, in the hope that this will lead to more realistic track shapes.

The structure of this paper is as follows.
First, in section~\ref{data} below, we describe the data set we use for this study.
Secondly, in section~\ref{properties}, we present various diagnostics of this data.
These diagnostics motivate the choice of a statistical model, which we describe in section~\ref{model}.
In sections~\ref{micro} and~\ref{macro} we then present results from this model,
at microscopic and macroscopic scales, respectively.
Finally in section~\ref{summary} we summarise our results.

\section{The data}
\label{data}

The basic data set for this study is a 54 year subset (1950-2003, inclusive) of the Hurdat data~\citep{hurdat}.
This 54 year period has 524 hurricanes, corresponding to 15,607 six-hourly observations of storm position.
In~\citet{hallj05a} we used a nearest neighbours method to estimate unconditional mean
hurricane tracks from this data. In~\citet{hallj05b} we then used a similar method to estimate the unconditional
variance of fluctuations around these mean tracks. In this study we again consider these fluctuations, but
now standardized using the variance model.
The 6-hour steps are projected onto the local optimized mean track. The mean track
(parallel and perpendicular components) are then subtracted, and the results divided by the
local optimized variance. Assuming the mean and variance models to be accurate, the result is 524 time series
of varying length, each with zero mean and unit variance.
We will generally refer to these standardized fluctuations
as `standardized anomalies' or just `anomalies'.

\section{Properties of the standardised anomalies}
\label{properties}
\begin{figure}[!htb]
  \begin{center}
    \scalebox{0.8}{\includegraphics{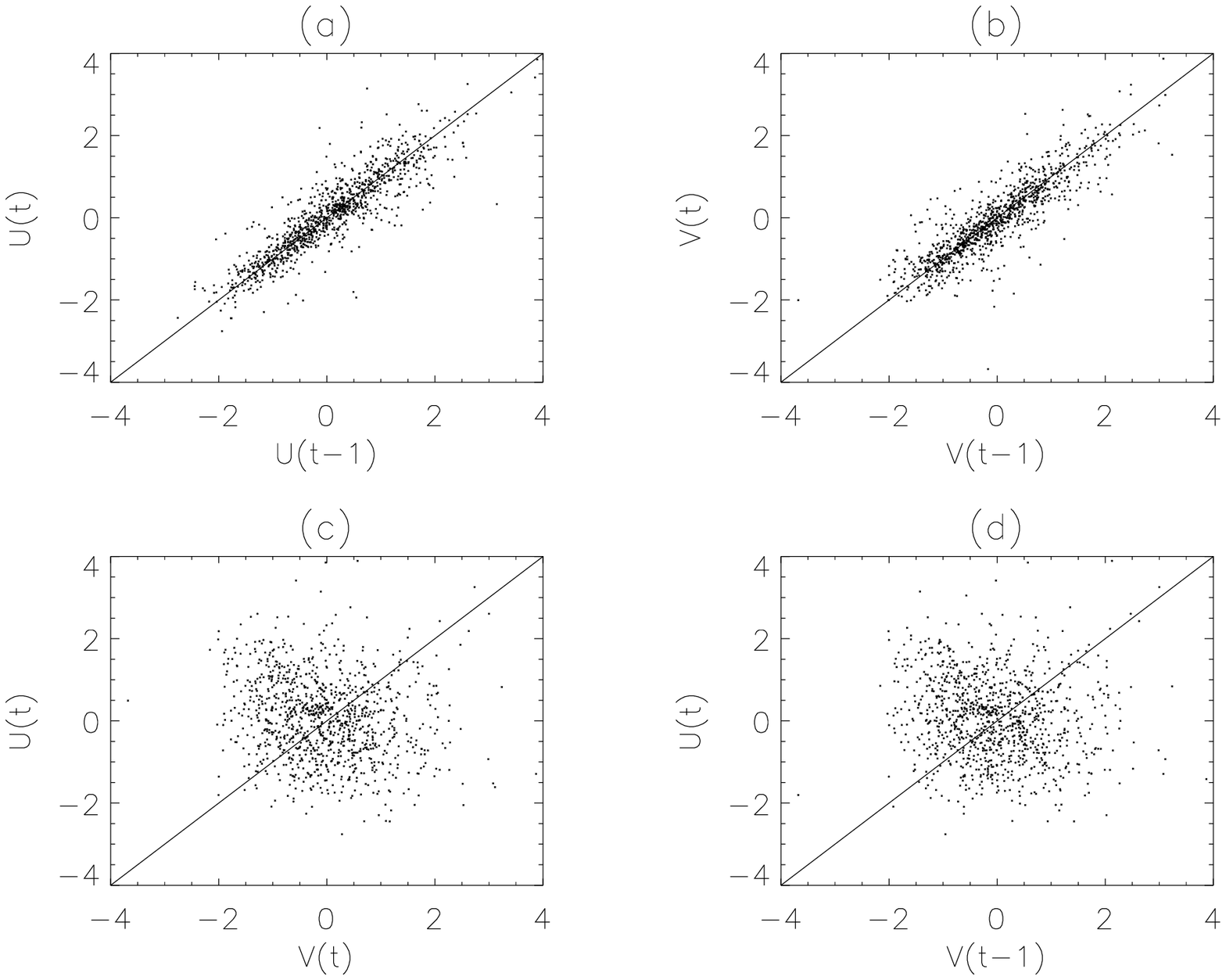}}
  \end{center}
  \caption{
Scatter plots that show some of the dependencies between anomalies.
Panel (a) shows the lag one dependency structure for along-track anomalies,
panel (b) shows the lag one dependency structure for across-track anomalies,
panel (c) shows the simultaneous dependency structure between along-track
and across-track anomalies and panel (d) shows the lag one dependency
structure between along-track and across-track anomalies, with across-track anomalies
leading.
     }
  \label{scatter}
\end{figure}

We now investigate the statistical properties of the standardized hurricane track anomalies,
beginning with the scatter plots in figure~\ref{scatter}.
Figure~\ref{scatter}a shows $u$ (i.e. along mean track) anomalies versus $u$ anomalies one step (6
hours) prior, and figure~\ref{scatter}b shows the same for $v$ (i.e. across mean track) anomalies.
Data are drawn from the full set of storm tracks, but only every 10th location is plotted for clarity.
In figure~\ref{scatter}c we plot $u$ versus $v$ at the same location,
while in figure~\ref{scatter}d $u$ is plotted versus $v$ at the previous step.
We see that $u$ and $v$ are closely and linearly related to their values at the previous step.
The presence of such a strong relationship in time is expected,
since hurricane motion is driven by large-scale weather patterns,
which themselves have memory.
By contrast, there is no evidence for a relationship between $u$ and $v$,
either contemporaneously or with a one-step lag, suggesting that independent modelling of
$u$ and $v$ would be appropriate.

\begin{figure}[!htb]
  \begin{center}
    \scalebox{0.8}{\includegraphics{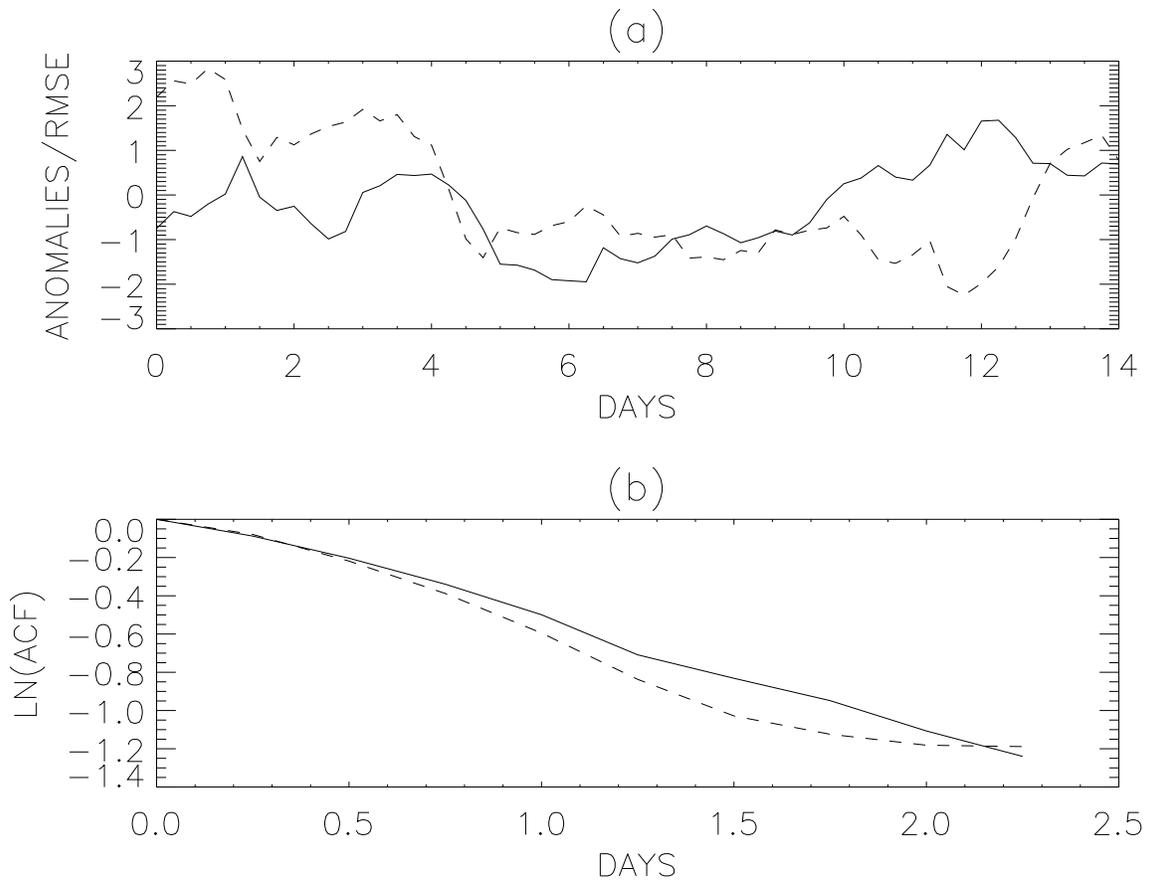}}
  \end{center}
  \caption{
The top panel shows a randomly chosen hurricane track  expressed in terms of the standardized fluctuations
around the unconditional mean track. The solid line shows the along-track fluctuations and the dashed line
shows the across track fluctuations.
The lower panel shows log-ACFs calculated from these tracks.
     }
  \label{series-example}
\end{figure}

To diagnose the nature of the $u$ and $v$ memory we compute autocorrelation functions (ACFs).
Figure~\ref{series-example} shows an example from a randomly selected storm track.
In figure~\ref{series-example}a we show the $u$ and $v$ anomaly time series for this track.
That there is memory is apparent in the lack of step-to-step variability.
Equally apparent is the lack of any obvious $u$-$v$ correlation.
The natural logs of the ACFs, shown in figure~\ref{series-example}b, display approximately linear decline.
A purely exponential  ACF is expected for a process that only carries information
forward from the previous time step (sometimes called a Markov process).

\begin{figure}[!htb]
  \begin{center}
    \scalebox{0.8}{\includegraphics{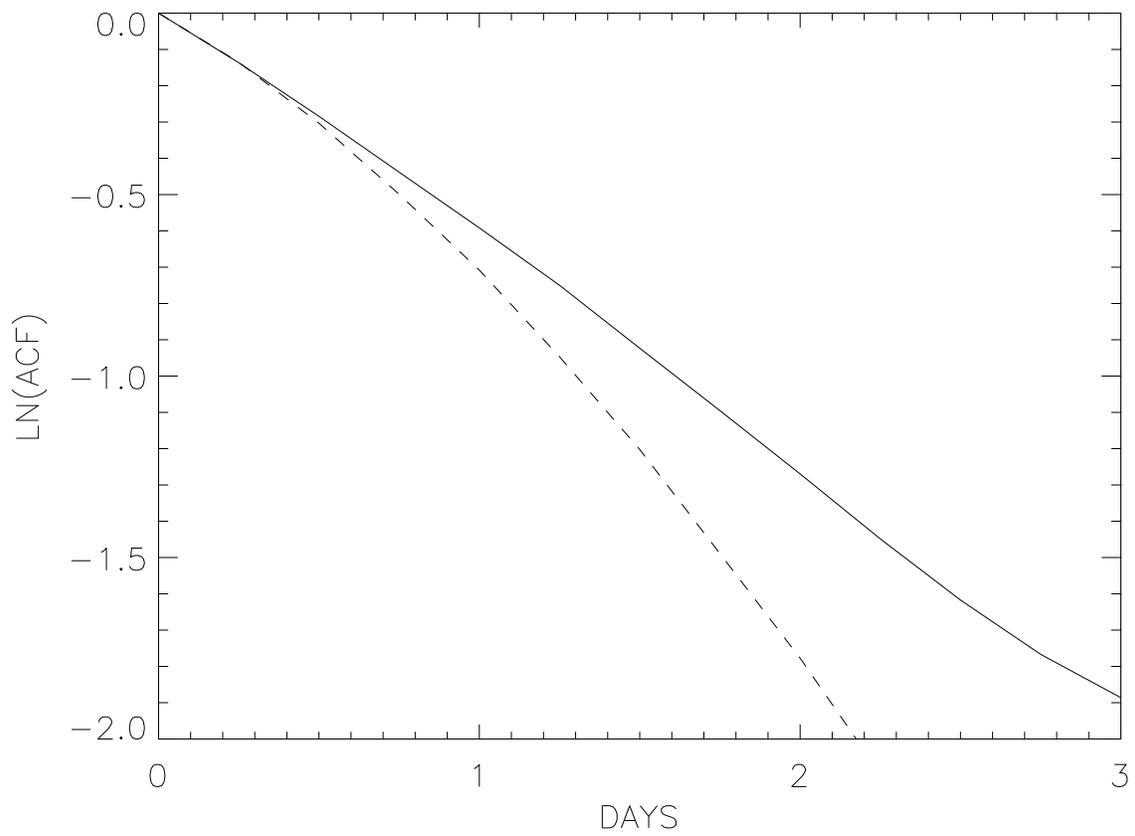}}
  \end{center}
  \caption{
Log-ACFs derived from the standard fluctuations from 524 storms over a 54 year period.
The solid line corresponds to along-track fluctuations while the dashed line corresponds to
across track fluctuations.
     }
  \label{acf}
\end{figure}

To test the robustness of these observations we also compute the ACFs
from the anomalies of all storms. These ACFs, up to a three-day lag, are shown in figure~\ref{acf}.
There is a large amount of data, and so these ACFs should be estimated well
(although the longer leads are based on fewer values than shorter leads because some storms last less than
3 days).  The $u$ log ACF is very close to linear.
The $v$ log ACF is slightly convex, indicating that $v$ anomalies have less memory than $u$ anomalies.

\begin{figure}[!htb]
  \begin{center}
    \scalebox{0.8}{\includegraphics{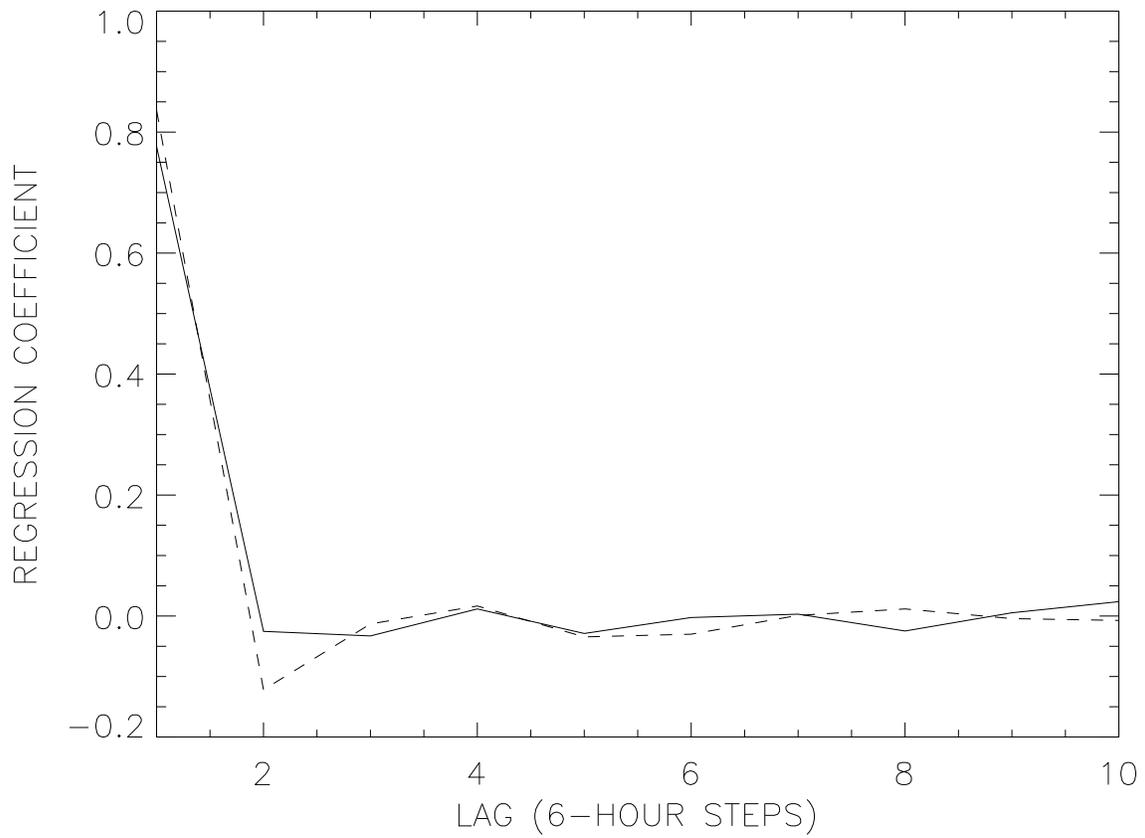}}
  \end{center}
  \caption{
Regression coefficients derived by regressing the standardised anomalies onto previous values.
The solid line corresponds to along-track fluctuations while the dashed line corresponds to
across track fluctuations.
     }
  \label{regression}
\end{figure}

In order to isolate the direct relationship between an anomaly and its value at various past
lags we perform a regression analysis
(ACFs are non-zero at lags greater than one even if the process is purely Markov
because of the lag-one correlations between intermediate steps).
Figure~\ref{regression} shows the coefficients obtained by a multiple regression of an anomaly
against the anomalies of the preceeding 10 steps on the same track, computed over all tracks of at least 10 steps.
At lag one the coefficients in $u$ and $v$ are about 0.8, but they are near zero at all subsequent lags.
We conclude from this analysis that a lag-one process considered separately for $u$ and $v$
is a reasonable model for the anomalies.

A lag-one ACF has the form $f(t) = e^{-t/\tau}$, where the autocorrelation time,
$\tau$, can be interpreted as the time scale for the track memory to decay.
We estimate this time scale in days from the observational ACFs using the value $f(t)$ at $t$ of lag one;
that is $t = 0.25$ days.  We plot $\tau$ as a function of space in figure~\ref{timescale-map}
using a 900km smoothing window
(the justification for which is presented in section 4 below).
There is spatial structure in $\tau$, with generally longer memory at higher latitudes.
Note that the memory in $u$ is somewhat larger than $v$, consistent with figure~\ref{acf}.

Finally, we ask to what degree the anomalies have a probability distribution that is normal
(in~\citet{hallj05b} we assumed a normal distribution when we derived the form of the likelihood that we
used to optimize the variance).
Figure~\ref{uqq-vqq} shows QQ plots of the $u$ and $v$ anomalies versus a normal distribution.
Were the observed anomalies purely normal, then the scatter would fall along a straight line.
In fact, the scatter only falls along a straight line inside $\pm$2 standard deviations.
Outside this range, the distribution of observed anomalies displays `fat tails', that is,
large anomalies are more common than would be the case if the distribution were normal.

In summary, we have investigated the local statistical properties of the observed hurricane track anomalies.
We find that the temporal dependence is simple: $u$ is linearly related to its value at the previous time step,
and likewise for $v$, but there is no evidence for a relationship to values at earlier time steps,
and no evidence for a relationship between $u$ and $v$ (note, however, that in
\citet{hallj05b} we saw weak, but significant,
$u$-$v$ correlations that varied in sign depending on geographic region).
The probability distributions of $u$ and $v$ are more complex.
They are normally distributed within $\pm$2 standard deviations, but beyond that show significantly fatter tails
than the normal distribution would.

\section{A linear autoregressive model}
\label{model}

Having investigated the local properties of the track anomalies, we now consider how they might be modelled.
The simplest standard statistical models for the modelling of
autocorrelated time-series are the autoregressive moving average (ARMA) models (see~\citet{boxj70}, or
almost any modern book on time-series analysis).
In the case of two time series, ARMA can be generalised to vector ARMA, which models
cross-correlations and cross-lag-correlations between the time-series. There are also other
generalisations of ARMA to include, for instance, long memory.
It would be convenient if we could use these standard models to capture the anomaly behaviour described above.
So we ask ourselves: based on the evidence we have seen so far,
are the ARMA models, or some generalisation of them, likely to be good for this purpose, and, if so,
which of the various ARMA models?

The lack of correlation between $u$ and $v$ suggests we need only consider separate univariate ARMA
models rather than a vector ARMA model that would model $u$ and $v$ as correlated.
The fact that the autoregression functions drop to zero after one lag suggests that we can simplify even
further and consider separate AR(1) models.
The linearity of the lag one dependency structure also suggests that using linear models such as AR(1) is reasonable.
The only catch seems to be that the QQ-plots show non-normality in the tails of the distribution, and the ARMA models
are based on an assumption of pure normality.
Given all of the above, we tentatively fit AR(1) models to the two time series,
bearing in mind that one of the
conditions for such a model to work (the distributional assumption) has not been completely satisfied.
The AR(1) models for $u$ and $v$ are then given by:

\begin{eqnarray}
 u_{n+1}&=&\phi^u_n u_{n} + s^u_n \epsilon^u_n\\
 v_{n+1}&=&\phi^v_n v_{n} + s^v_n \epsilon^v_n
\end{eqnarray}

There are two coefficients in each equation: the memory parameter $\phi_n$
and the standard deviation of the noise forcing $s_n$. Since our anomalies are already
standardised, there is a simple constraint between these two parameters.
This constraint, and other details of the model, are described in detail in appendix~\ref{ar1model}.

\begin{figure}[!htb]
  \begin{center}
    \scalebox{0.8}{\includegraphics{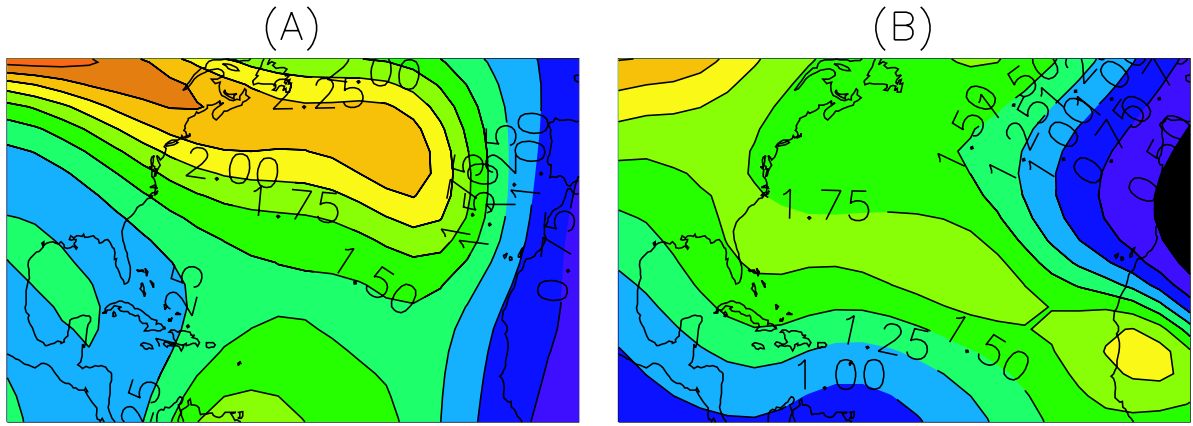}}
  \end{center}
  \caption{
The lag one correlation for along-track and across-track anomalies, converted into a timescale
with units of days.
     }
  \label{timescale-map}
\end{figure}

\begin{figure}[!htb]
  \begin{center}
    \scalebox{0.8}{\includegraphics{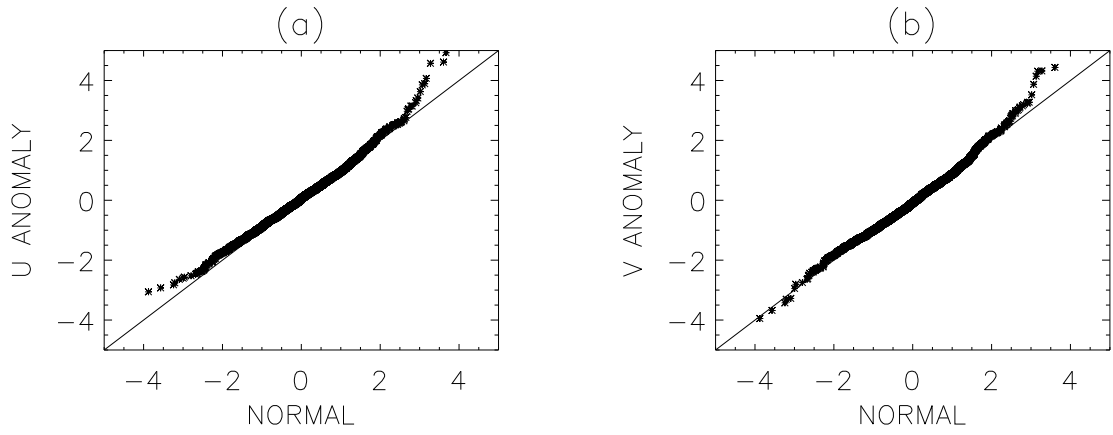}}
  \end{center}
  \caption{
QQ plots comparing the distributions of along-track and across-track anomalies with the normal distribution.
     }
  \label{uqq-vqq}
\end{figure}

\begin{figure}[!htb]
  \begin{center}
    \scalebox{0.8}{\includegraphics{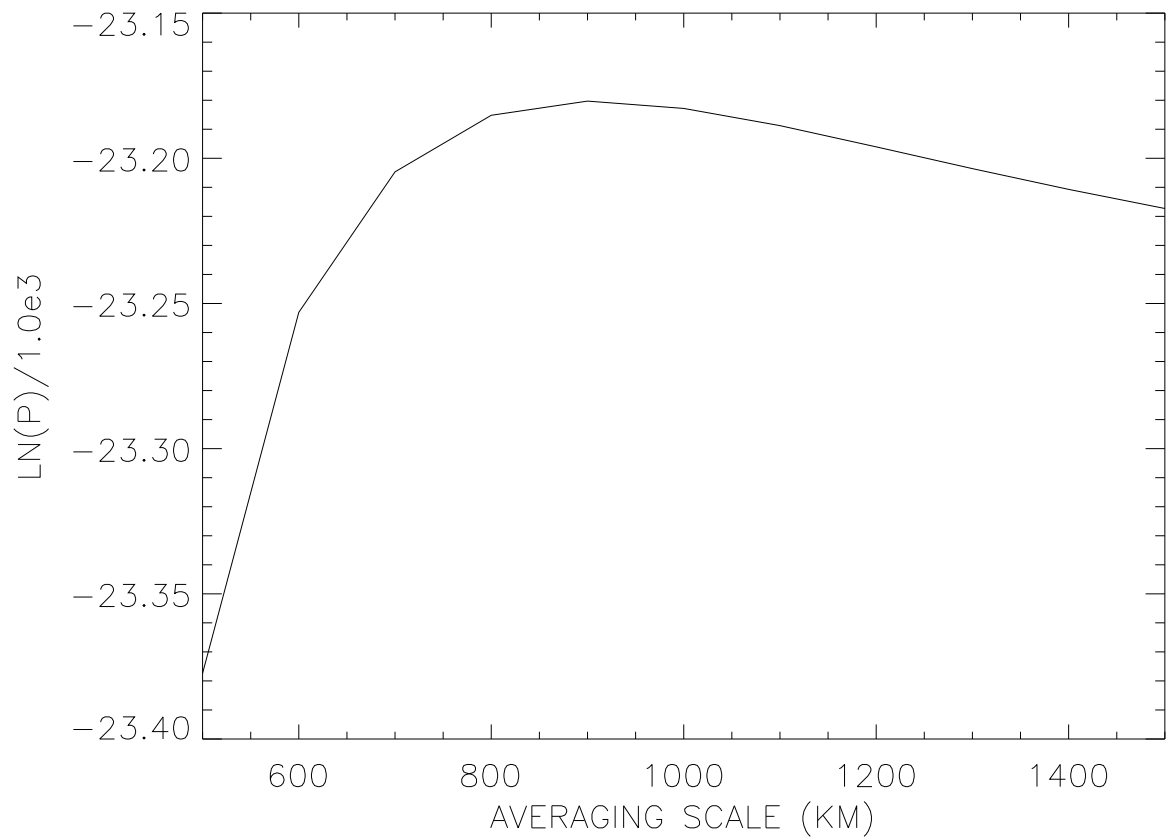}}
  \end{center}
  \caption{
The out-of-sample log-likelihood versus the smoothing length-scale used to estimate the
lag one autocorrelation, showing an optimal lengthscale of 900 km.
     }
  \label{scale}
\end{figure}

The simplest implementation of this model would have the memory parameter
and noise variance constant in space.
However, as we have seen above in figure~\ref{timescale-map},
this would not be entirely realistic since the autocorrelation
time-scale of the observations clearly varies. We therefore allow both the memory parameter
and the noise variance to vary in the model,
and in fact the value of $\phi$
(and the value of $s$, which is derived from $\phi$)
is derived directly from the data used to create figure~\ref{timescale-map}.
The spatial smoothing used to create $\phi$ was derived optimally using the same
method as used to define the optimal smoothing in the modelling of the mean
and the variance: each year of the 54 years of historical data was eliminated
in turn, and for a range of values of the smoothing lengthscale the log-likelihood
(given by equation~\ref{ll} in appendix~\ref{ar1model}) for the eliminated year was calculated
using the AR(1) model fitted to the rest of the data.
The smoothing window has a Gaussian shape, and is normalized so that the weights
sum to one.
The resulting dependence of log-likelihood on
lengthscale, which gives an optimum lengthscale of 900 km, is shown in figure~\ref{scale}.

How well does the AR(1) model work? There are many tests that we can do to try and answer this question, and
the rest of this article consists of results from a variety of such tests, and
some discussion and speculation about what the results might be telling us.

\section{Microscopic diagnostics}
\label{micro}

We will start our assessment of the performance of the AR(1) model using what we call `microscopic'
diagnostics. By this we mean an assessment of how well the model reproduces the statistics of individual
steps of hurricane tracks. We will then continue our assessment (in section~\ref{macro} below)
by looking at `macroscopic' diagnostics,
by which we mean an assessment of how well simulated tracks from
the model reproduce the large scale features of the observed hurricane
track distribution.

\begin{figure}[!htb]
  \begin{center}
    \scalebox{0.8}{\includegraphics{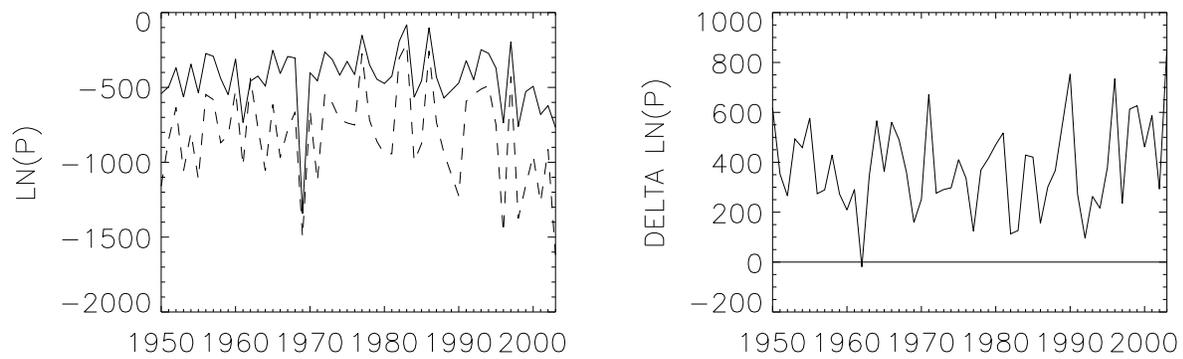}}
  \end{center}
  \caption{
The left panel shows out-of-sample log-likelihood scores for AR(0) (dashed) and AR(1) (solid) models.
The right panel shows the difference between them.
     }
  \label{yearly}
\end{figure}

In our earlier articles we discussed the importance of performing model selection
using out-of-sample testing, and we implemented such testing using the Quenouille-Tukey jack-knife.
Our first evaluation of the AR(1) model is to apply such testing and compare the AR(1) model to
the simpler AR(0) model. The AR(0) model neglects the memory in $u$ and $v$ entirely and models them
both as stationary `white noise'.
Thus comparing AR(0) with AR(1) is a test for the importance of modelling the memory.
The comparison between the AR(0) and AR(1) models is performed by missing out each
year of the data in turn, fitting the model to the rest of the data, and evaluating the likelihood
of the missed year for both models.
The likelihoods for all missed years are then compared across the two models on a year
by year basis. We like this test because it directly compares the ability of the models to extrapolate
the distribution of possible hurricanes,
and extrapolation of the hurricane distribution is exactly what is needed
for good assessments of hurricane risk.
The year by year comparison of out of sample likelihoods is shown in figure~\ref{yearly}
(the left panel shows log-likelihood scores for the two models
and the right panel shows the differences between them).
We see a dramatic difference between the two models: the AR(1) model beats AR(0) in every year except
one. If the models were in fact equally good this could only occur with extremely low probability,
and so this seems to be conclusive proof that, by this measure, modelling the correlations in the track
fluctuations gives an improvement over not modelling them. This is not entirely surprising, of course.

Our second set of microscopic diagnostics concerns the behaviour of the residuals from the model.
If the model were completely correct then the residuals would be uncorrelated in time,
normally distributed and not cross-correlated.
On the other hand, if any of these properties do not hold, that immediately gives
an indication that the model is not ideal. And since we already know that the original anomaly data
didn't perfectly satisfy the assumptions behind the AR(1) model, we might already anticipate some problems with
the residuals too.

\begin{figure}[!htb]
  \begin{center}
    \scalebox{0.8}{\includegraphics{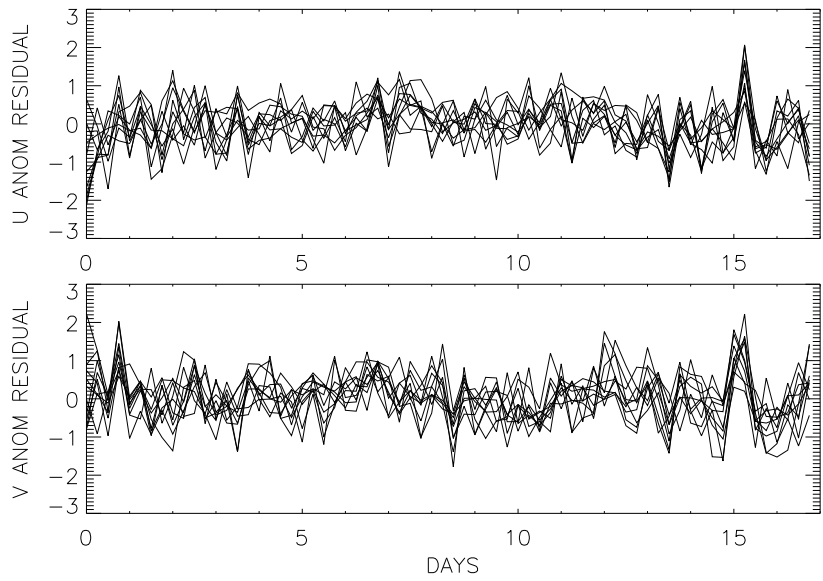}}
  \end{center}
  \caption{
Several examples of time series of residual anomalies in the along-track and across-track directions.
     }
  \label{residual-anomaly-series}
\end{figure}

\begin{figure}[!htb]
  \begin{center}
    \scalebox{0.8}{\includegraphics{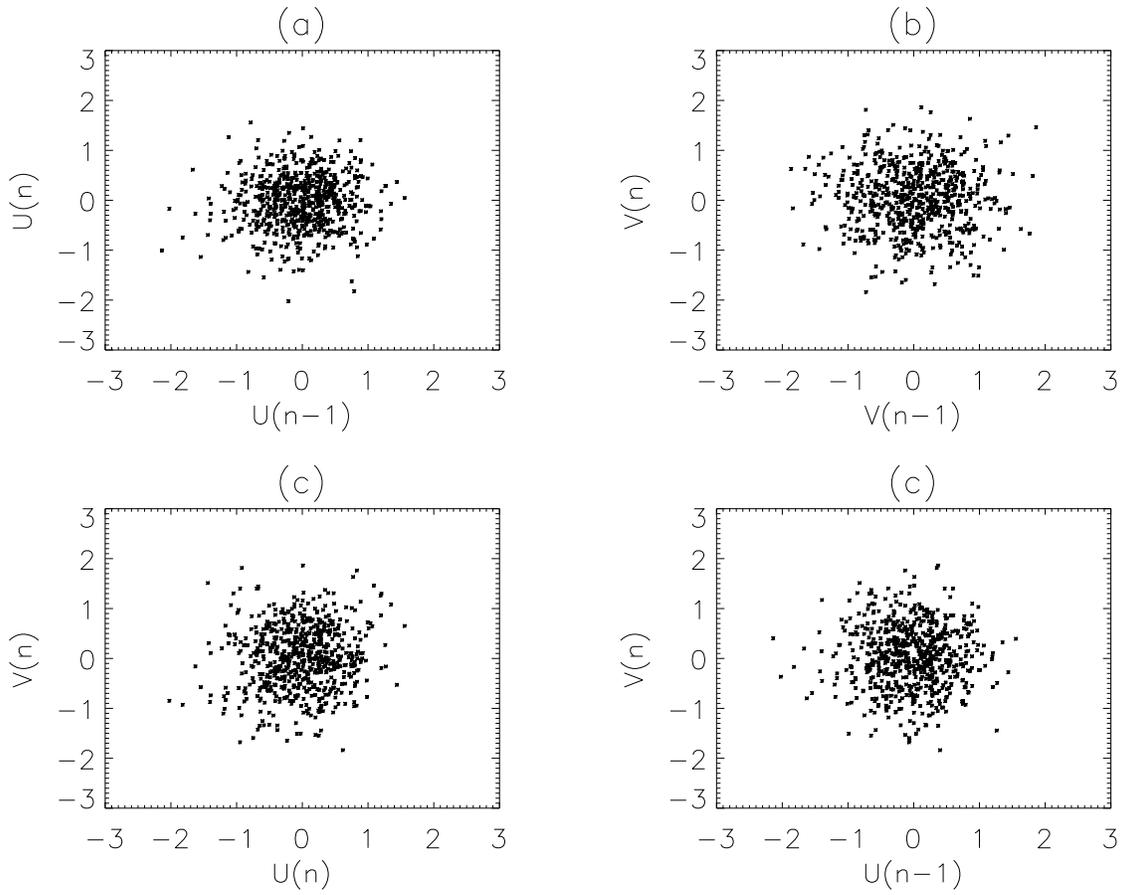}}
  \end{center}
  \caption{
Scatter plots that show some of the dependencies between anomaly residuals.
Panel (a) shows the lag one dependency structure for along-track residuals,
panel (b) shows the lag one dependency structure for across-track residuals,
panel (c) shows the simultaneous dependency structure between along-track
and across-track residuals and panel (d) shows the lag one dependency
structure between along-track and across-track residuals, with across-track residuals
leading.
     }
  \label{residual-anomaly-scatter}
\end{figure}

We start our residual analysis by showing a number of superimposed time series of residual anomalies
from different storms,
in figure~\ref{residual-anomaly-series}.
Various lag and cross-correlations derived from the residual anomalies
of all storms are then shown in figure~\ref{residual-anomaly-scatter}
(note that this figure looks very similar to figure~\ref{scatter}, but is now for residuals rather than
the anomalies themselves).
The first panel shows the lag one
dependence for $u$, the second panel shows the lag one dependence for $v$, the third panel shows the
lag zero dependence between $u$ and $v$ and the fourth panel shows the lag one dependence
between $u$ and $v$, with $u$ leading. In all four cases the results are very good and there is no
discernible structure. This suggests that the model is doing a good
job of capturing the behaviour of the anomalies in time, and this accords well with the analysis of the
anomalies themselves, discussed in section~\ref{properties} above,
in which we saw that the regression coefficients dropped to zero after one lag and the relation
between adjacent anomalies was close to linear
(both of which suggest that AR(1) may work well).

\begin{figure}[!htb]
  \begin{center}
    \scalebox{0.8}{\includegraphics{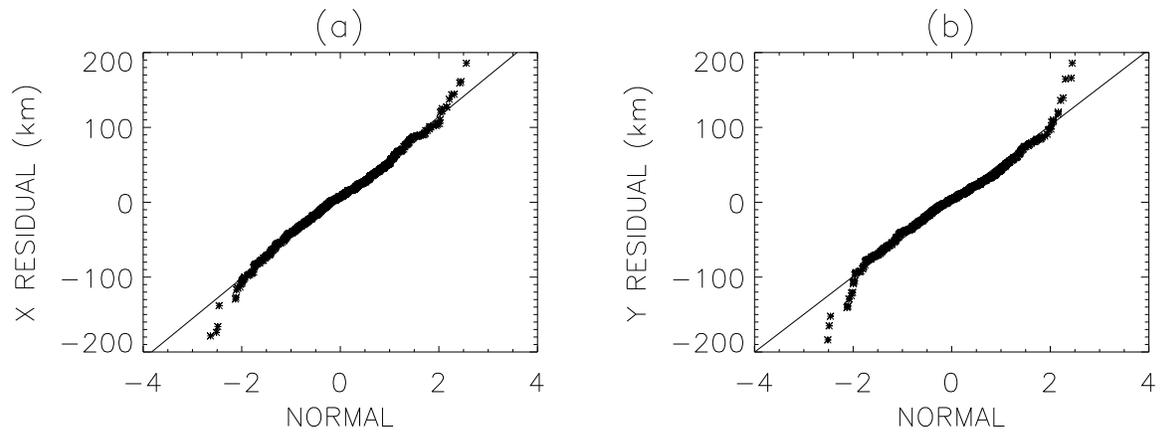}}
  \end{center}
  \caption{
QQ plots comparing the distributions of the residual anomalies with the normal distribution.
     }
  \label{residual-52-QQ}
\end{figure}

The distributions of the $u$ and $v$ residuals are shown in figure~\ref{residual-52-QQ}.
The horizontal axes are shown as quantiles while the vertical axis is in km.
We see that the residuals are close to normal out to about 2 standard deviations, but then deviate sharply
from normal, and show definite fat tails. This is very similar to the behaviour of the anomalies themselves, and
is a clear fault with the model, since the residuals should be normally distributed.
It would seem likely that the fat tails of the residuals are a direct consequence of the fat tails in the
original data.
What might be causing these fat tails?
Possibly that a small fraction of storms behave very differently to the majority.
This could be storms in certain seasons, or certain regions, or in certain large-scale weather conditions.
Or maybe it is wrong to look for a particular cause: why, after all, should the distributions of the
anomalies and the residuals be normal in the first place?

Based on all of the microscopic diagnostics discussed above
we conclude that our model captures some, but not all, of the
observed behaviour of the motion of hurricanes at a microscopic level.
The temporal dependence of the tracks seems to be captured very well.
The main shortcoming would seem to be the
inability of the model to capture the correct distribution of the anomalies, which is fatter than normal in
the tails.
At a statistical level there is no single obvious way to remedy this distribution problem.
Two of the things one might consider trying would be:

\begin{itemize}

    \item using non-normal residuals in the subsequent simulations,
    either modelled or resampled from the residuals generated during the fitting process.

    \item transforming the anomalies to normal before fitting the AR(1) model, and transforming them back
    to the original distribution.

\end{itemize}

However, rather than just blindly testing these, or other, statistical methods we will first consider
the macroscopic diagnostics. This may lead to some more `physical' insight into why the model is not
capturing the correct distribution.

\section{Macroscopic diagnostics}
\label{macro}

We now move on to consider macroscopic diagnostics from our fitted track model.
These are diagnostics that assess whether the tracks that can be generated from the track model
are realistic or not. If the microscopic diagnostics indicated that the model were perfect,
then we would expect the tracks to be realistic. However, we have seen that the
microscopic diagnostics are not perfect, and that there are systematic biases in the tails
of the distribution of simulated track steps. Furthermore, there may be other biases
that we haven't detected (for instance, we haven't considered the microscopic diagnostics
on a region by region or month by month basis, and this could be obscuring problems in the model).
It is very hard to know how important any microscopic
biases are likely to be when it comes to simulating full tracks, and whether they will somehow
cancel out, or whether they will lead to large systematic errors in the track shapes.
All of the following diagnostics are therefore based on integrating the track
model to generate simulated tracks, and comparing these tracks with the observed.

\begin{figure}[!htb]
  \begin{center}
    \scalebox{0.8}{\includegraphics{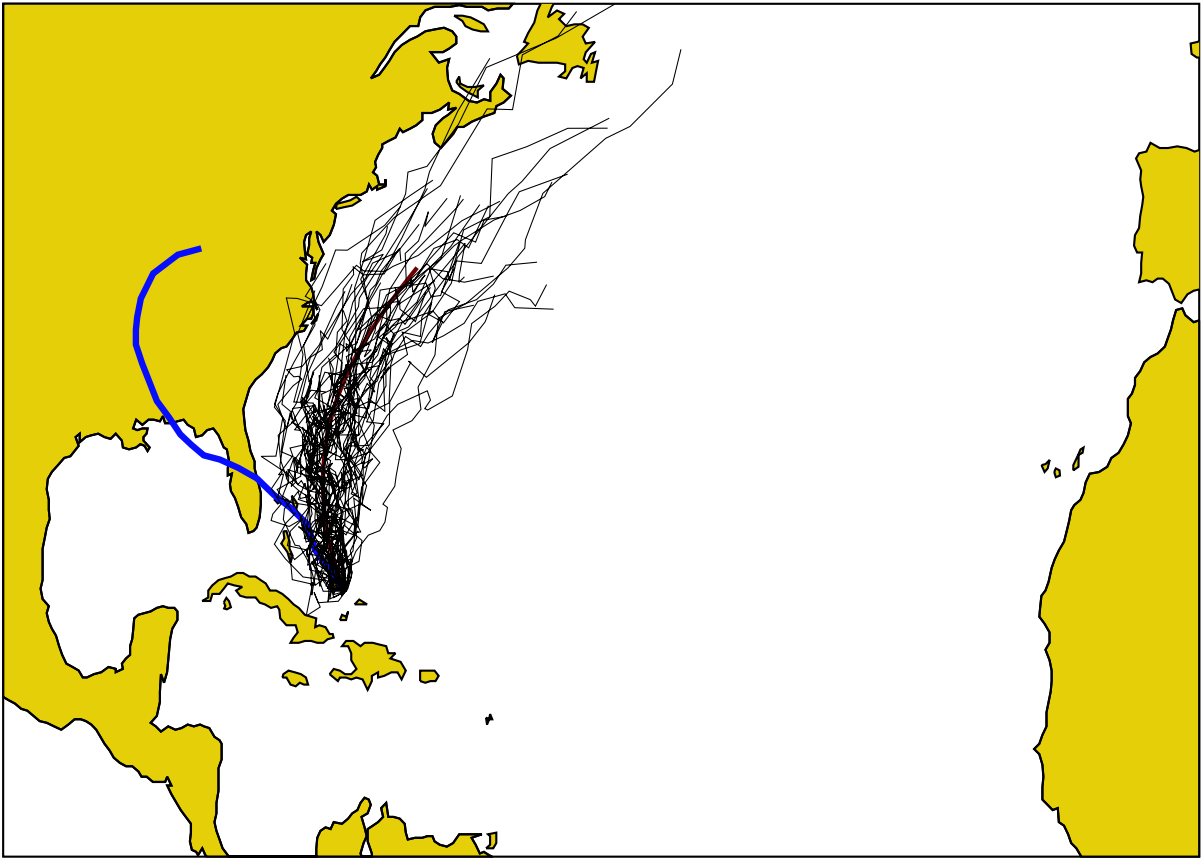}}
  \end{center}
  \caption{
For a fixed origin point, the red line shows the unconditional mean track, the blue line shows a real track
and the black lines show simulated tracks from an AR(0) model.
     }
  \label{tracks-no-mem}
\end{figure}

\begin{figure}[!htb]
  \begin{center}
    \scalebox{0.8}{\includegraphics{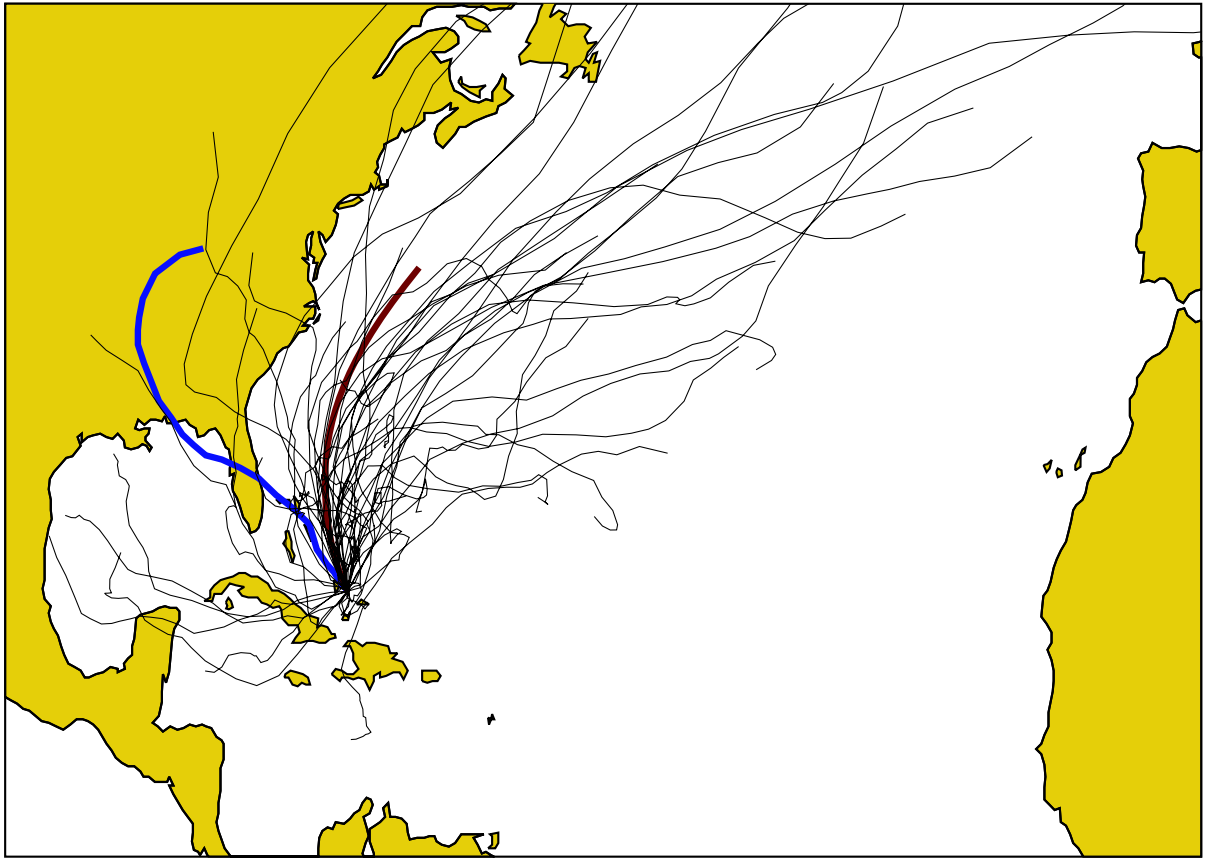}}
  \end{center}
  \caption{
As for the previous figure, but now for simulations with an AR(1) model.  }
  \label{tracks-mem}
\end{figure}

Our first comparison is based on the simulation of tracks from a single point with the AR(0) and
AR(1) models (see figures~\ref{tracks-no-mem} and~\ref{tracks-mem}).
The red lines
in both cases show the mean track from that point, the blue lines shows a real track
and the black lines show a number of simulated tracks.
The AR(0) model (figure~\ref{tracks-no-mem}) gives tracks that are rather
similar to each other and very bunched together. It is clear that the real track is qualitatively different
from the simulated tracks, and hence that the simulated tracks do not come from an appropriate model.
The AR(1) model (figure~\ref{tracks-mem})
performs much better: the simulated tracks are now very different from each other, and spread out
rapidly. It now seems quite plausible that the real track could have been generated as one of these simulations,
and hence by this test the AR(1) model is a good one. One interesting point in this comparison is that the
AR(0) and AR(1) models have the same variance of the anomalies: all that differs is the autocorrelation.
We thus see that the rapid spreading of the AR(1) tracks is due entirely to this autocorrelation.

\begin{figure}[!htb]
  \begin{center}
    \scalebox{0.8}{\includegraphics{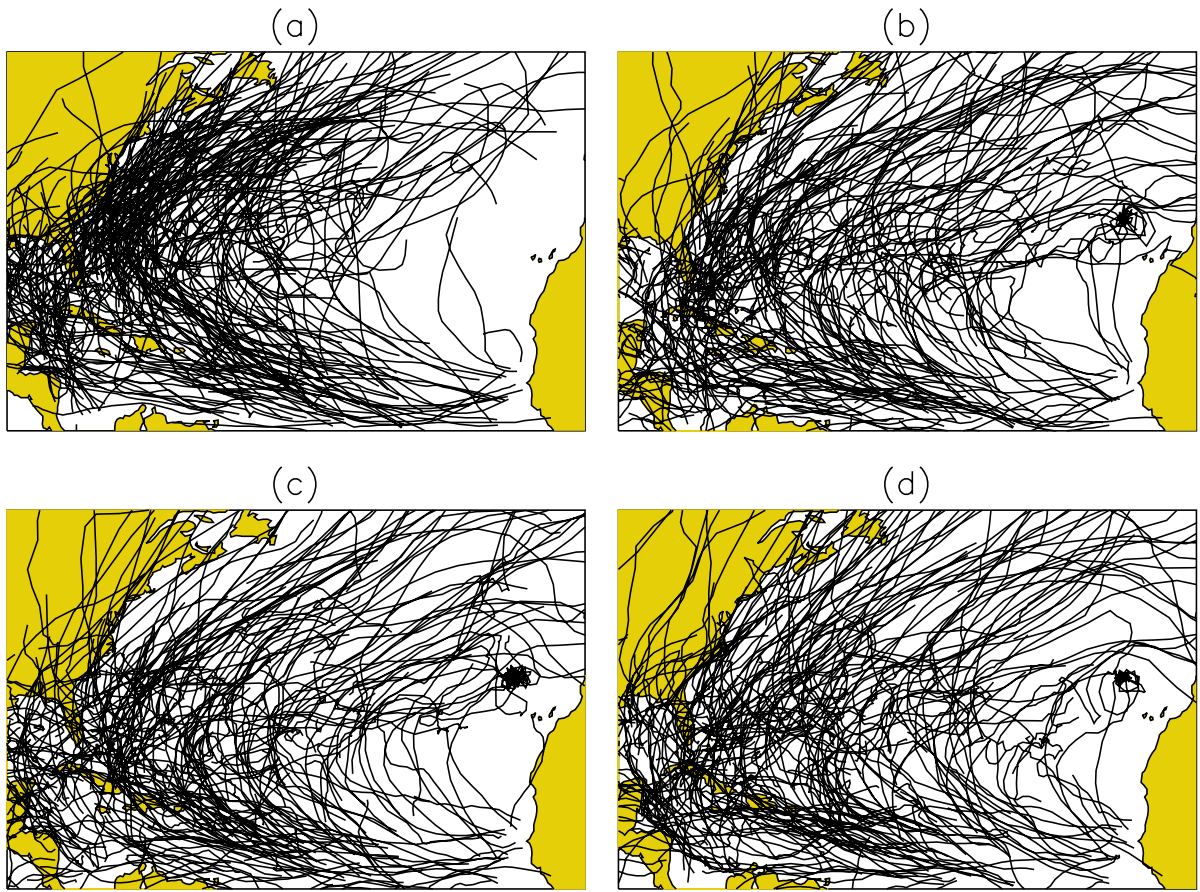}}
  \end{center}
    \caption{
Panel (a) shows historical hurricane tracks for a period of 54 years.
The other panels show simulated hurricane tracks, starting at the same origin points as the historical
tracks, generated using the AR(1) model.
     }
  \label{tracks}
\end{figure}

\begin{figure}[!htb]
  \begin{center}
    \scalebox{0.8}{\includegraphics{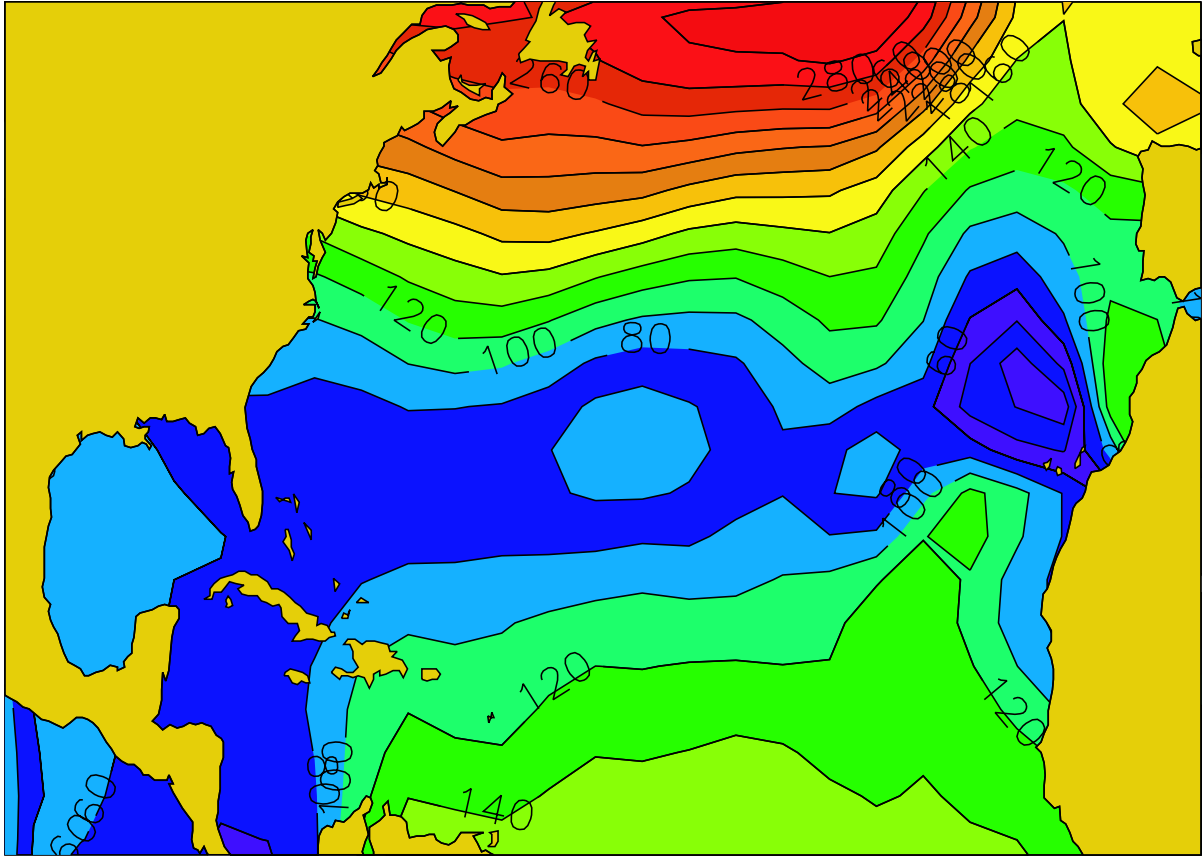}}
  \end{center}
  \caption{
The forward speed in the mean track model, in units of km per 6-hour time step.
     }
  \label{speed}
\end{figure}

So far we have seen that the AR(1) model performs well in comparison with the simpler AR(0) model.
But AR(0) is an easy target to beat, and we do not yet have any idea how the AR(1) model performs
in an absolute sense. Figure~\ref{tracks} is an attempt to look at that question. It shows
the 524 observed tracks for our historical data set (in panel (a)), and three realisations of
the same number of simulated tracks (in the other three panels).
The simulated tracks were generated from the same starting points as the observed tracks,
and each track has the same number of steps as the corresponding real track.
We shouldn't expect the simulated tracks to look exactly like the observations.
But if the AR(1) model were a perfect model, and if track origin and track shape
are independent, then it should look as if the observations could have
come from another realisation of the model.
The first thing we note from making this comparison
is that the simulations show a singularity of some sort in the eastern Atlantic. The cause
for this becomes rather clear when we consider the forward speed from the mean
track model (figure~\ref{speed}). The forward speed in this region is very small compared to
the forward speeds elsewhere in the basin: any simulated track that enters this region will likely take
a long time to leave it. Also, comparison with the observed tracks shows that there are no tracks in this region
at all. We conclude that the singularity is therefore just an artefact of the model, for two reasons.
Firstly, we are
running a track model but no simulation of the hurricane intensity. In reality, any hurricane
entering this region of the Atlantic would die because the sea surface temperatures are too low to
support the evaporation needed to drive the hurricane. Thus the singularity would likely disappear if we were to add
an appropriate intensity model (which we intend to do in due course).
Secondly, our mean track speeds are given by the average observed speeds of
nearby hurricane tracks. In regions where there are no tracks this average takes values from far away
and ceases to be physically meaningful. We could equally well set the mean track speed in this region to some
arbitrary (but reasonable) fixed value, and this would eliminate the singularity.

\begin{figure}[!htb]
  \begin{center}
    \scalebox{0.8}{\includegraphics{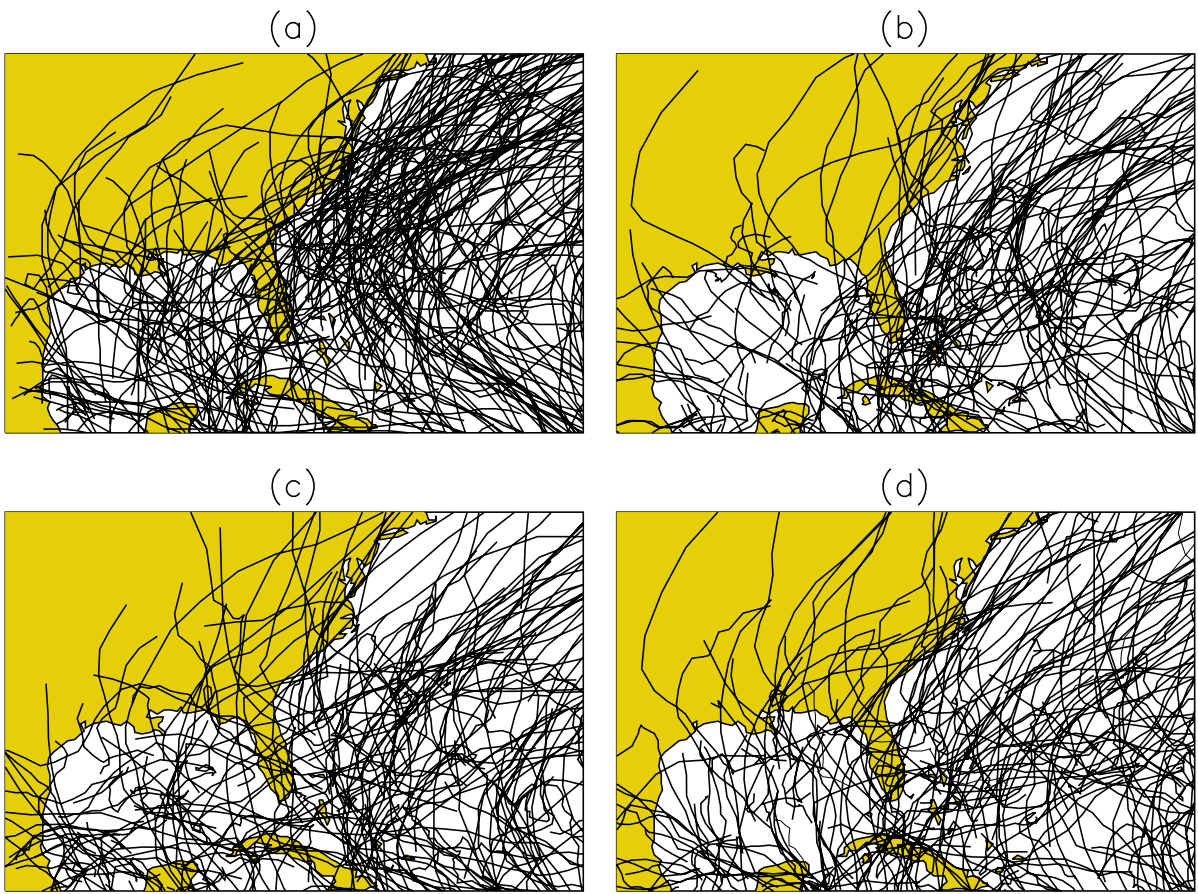}}
  \end{center}
  \caption{
  Same as figure~\ref{tracks} but for a smaller region.
  }
  \label{tracks2}
\end{figure}

Looking more generally at the shapes of the simulated tracks we see that
overall the tracks from the model seem reasonably similar to the observed tracks,
and certainly capture the broad behaviour of the observations. However, in detail, there
are differences.
Too many tracks penetrate deep into the America continent, and too many cross central America
into the Pacific. These two faults would probably be solved by the inclusion of an intensity model
that would presumably kill these tracks a little earlier.
More seriously, perhaps, the density of tracks off the east coast of the US is
higher in the observations than any of the sets of simulations.
This is highlighted in figure~\ref{tracks2}, which shows a close-up of the US coastline,
with the observed and simulated track sets as before.

\begin{figure}[!htb]
  \begin{center}
    \scalebox{0.8}{\includegraphics{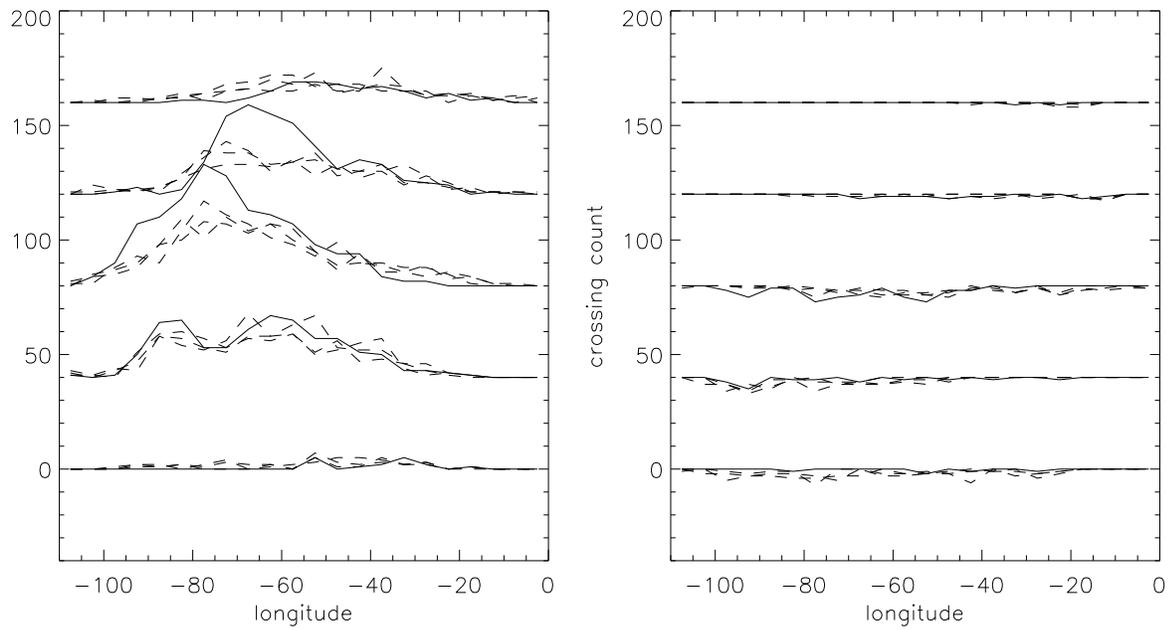}}
  \end{center}
  \caption{
  The number of tracks from observations and from simulations that cross certain lines
  of latitude (equally spaced from 10N to 50N, from bottom to top), in a northward
  direction (left panel) and in a southward direction (right panel).
  }
  \label{latcross}
\end{figure}

Figure~\ref{latcross} considers in more detail the number of observed and simulated tracks
that cross individual lines of latitude moving from South to North (left panel) and
North to South (right panel). The various sets of lines correspond to different latitudes
(which are, from bottom to top, 10N, 20N, 30N, 40N and 50N),
the bold line shows the numbers of hurricanes in 54 years of observations and the dashed lines
show numbers of hurricanes from 3 sets of simulations.
If the model were perfectly correct than the observations should lie
close to the sets of simulations. For the northward moving storms the model
seems to perform well at 10N and 20N, but at 30N and 80W there is a greater number
of observed storms than simulated. This is also the case at 40N and 70W. This corresponds
exactly to the region off the US east coast in figure~\ref{tracks2} where the simulations do
not generate a great enough density of storm tracks.

\begin{figure}[!htb]
  \begin{center}
    \scalebox{0.8}{\includegraphics{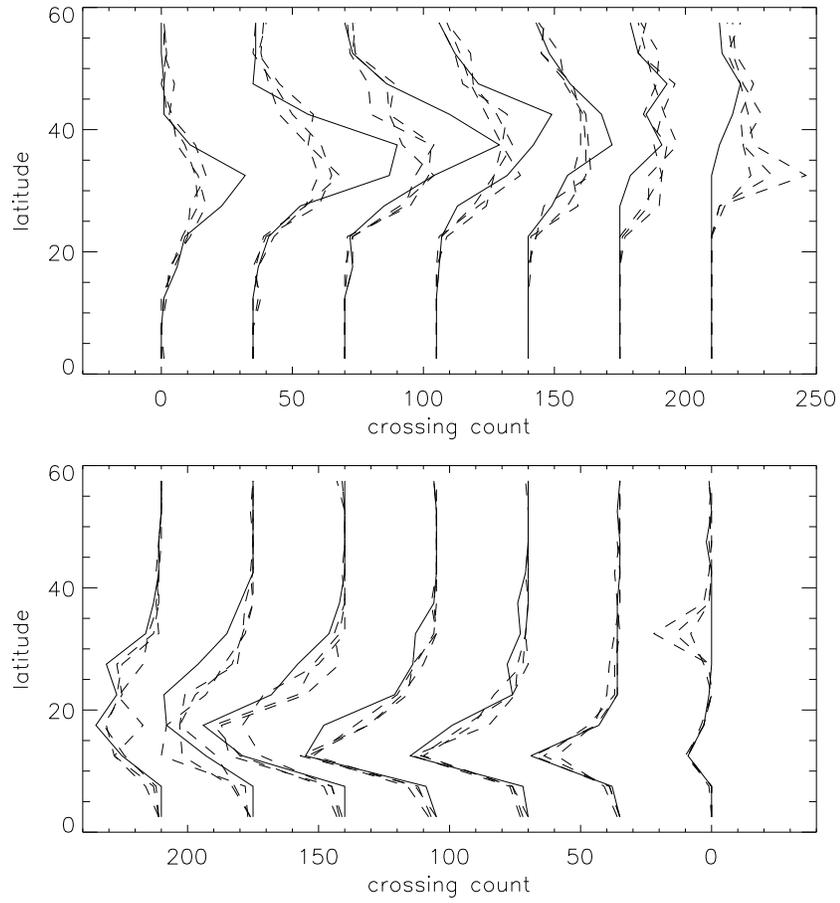}}
  \end{center}
  \caption
  {
  The number of tracks from observations and from simulations that cross certain lines
  of longitude (equally spaced from 80W to 20W, from left to right), in a eastward
  direction (panel (a)) and in a westward direction (panel (b)).}
  \label{loncross}
\end{figure}

Figure~\ref{loncross} counts observed and simulated storms moving from West to East (upper panel)
and from East to West (lower panel) in a similar way. The longitudes used are
80W, 70W, 60W, 50W, 40W, 30W and 20W (from left to right).
The model does reasonably well for the storms moving
East to West in the subtropics, but less well for storms moving from West to East in the extratropics.
In particular, there are too few storms moving from West to East at longitudes of 80W, 70W, 60W and
50W, and for latitudes between 30N and 45N.

\begin{figure}[!htb]
  \begin{center}
    \scalebox{0.8}{\includegraphics{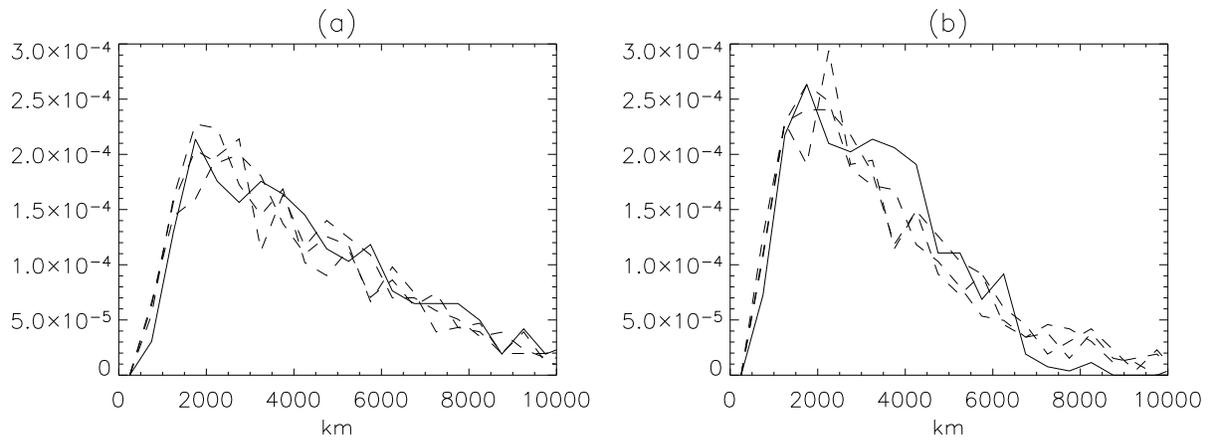}}
  \end{center}
  \caption{
  Panel (a) shows the distributions of track lengths for model and observations.
  Panel (b) show the distributions of distances from the start of each track to the end of the
  track for model and observations.}
  \label{pdfs}
\end{figure}

Our next diagnostic is designed to test whether the speed of the simulated tracks is correct.
We test this in two ways, both shown in figure~\ref{pdfs}.
The left hand panel shows the distribution of the total length of the trajectories: model and observations
agree very well. The right hand panel shows great circle distance from the start to end
of each trajectory: again, model and observations show fairly good agreement, although there are
more simulated tracks than observed tracks that reach distances greater than 7000km from their origins
(again, this may be related to the lack of a physical death mechanism, since some storms reach
unrealistically high latitudes).

Our final diagnostic is an attempt to create a measure of track density in space
i.e. to count how many storms are passing through different parts of the basin. We calculate this
measure as follows. First, we divide the whole basin into 100km-by-100km grid cells. Within
each grid cell we then count the number of `storm points', which are six-hourly locations of storms.
Slow moving storms may be counted several times in a single box.
We plot both the observations, and results from an ensemble of 34 simulations, each of which
contains 54 years of simulated tracks starting from the observed genesis points as before.
This allows us to compare the observed track density against the distribution of results
generated by the track model.
Panel (a) is the number of simulated storm points per 100km-by-100km box accumulated over 54 years and
averaged over the ensemble.  Panel (b) is the historical distribution of the same quantity, and panel
(c) is the difference between the historical and simulated.
We see immediately that the model is failing to create the very high density of observed
storms off the US coast that is seen in the observations, exactly as discussed above.
To be rigourous, however, one might question whether these differences
are significant (they certainly \emph{look} significant, but we should test to be sure).
To this end, panel (d) is the standard deviation across the simulation ensemble members.
Comparison of this with panel (c) makes it clear that the differences between the observed
and simulated fields are 5 or 6 standard deviations of the distribution of the simulated field,
which confirms that these are significant
differences, and not just due to randomness.

\section{Summary}
\label{summary}

We have described the third stage of our attempt to build a statistical model that gives faithful simulations
of the tracks of Atlantic hurricanes. The first stage was to model the mean, the second stage was to model the
variance, and the third stage, described above, has been to attempt to model the autocorrelations.
We have shown that modelling the autocorrelations allows us,
for the first time in this study, to create simulations of hurricane tracks that look somewhat realistic.

Detailed analysis of the residuals from the model shows that we are capturing the temporal dependency
structure of hurricane track anomalies very well, but that the distribution of track anomalies is fat-tailed
and hence that our model, which assumes Gaussian distributions, is deficient in this respect.

Comparison of simulations of tracks from the model with observed tracks show that the model is broadly correct, but
has a number of systematic errors when analysed in detail.
Some, but not all, of these systematic errors would be eliminated if we included an intensity model in our simulations.
In particular, the observations show a very high density of tracks off the East Coast of the US, and this
is not replicated by the model. This shows up in a number of diagnostics such as simulated track sets,
counts of storms crossing longitude and latitude lines and storm point density plots.

The next stage in our research will be to try and better understand, and hopefully correct, this systematic
bias. There are many candidates for conditioning the calculations of the mean, variance and autocorrelation,
such as time of year, storm intensity, geographic region of storm origin, and climate indices such as
ENSO and the North Atlantic Oscillation. Also, including some representation of the `fat tails' of the anomaly
probability distribution would certainly improve the microscopic performance of the simulations, and,
possibly, the macroscopic performance.

Clearly, a physical death process is needed to stop the propagation of simulated storms in
meteorologically unfavourable regions. We believe, however, that the infrastructure we have developed,
including the out-of-sample validation and the micro- and macroscopic diagnostics, will allow us to
assess readily if new features results in significant improvement.

\appendix
\section{The AR(1) model}
\label{ar1model}

\subsection{Likelihood for AR(1)}

Consider an observed hurricane track with $m+1$ points.
To define the likelihood for AR(1)
we need to use an $m$ dimensional normal distribution
($m$ not $m+1$ because the first point is considered fixed, and only
subsequent points are random).

We consider the $u$ component ($v$ is analogous).

The AR(1) model has the form:
\begin{equation}
\label{ar1u}
 u_{n+1}=\phi_n u_{n} + s^u_n \epsilon_n
\end{equation}

where $s_n$ is the standard deviation of the noise forcing
(normally one would use $\sigma$ rather than $s$, but we've used $\sigma$ already in equation 1 in~\citet{hallj05b}).

Writing out the first few terms of the model:

\begin{eqnarray}
  u_0 &=& 0 \mbox{    (starting point of the track)} \\
  u_1 &=& s_0^u \epsilon_0 \\
  u_2 &=& \phi_1 u_1+s_1^u \epsilon_1 \\
  u_3 &=& \phi_2 u_2+s_2^u \epsilon_2 \\
  ... &=& ... \\
  u_m &=& \phi_{m-1} u_{m-1}+s_{m-1}^u \epsilon_{m-1}
\end{eqnarray}

It looks like there are two parameters ($\phi$ and $s$), but really there is only one.
This is because if we square equation~\ref{ar1u},
take expectations, and use the fact that the variance of $u$ is 1
(because we have used the variance model to standardize the anomalies)
then we find that
\begin{equation}
 (1-\phi^2)=\sigma^2
\end{equation}

So all we have to do is to estimate $\phi_n$ and we can calculate the amplitude of the noise, $s_n$.

$\phi_n$ is the correlation between $u_{n+1}$ and
$u_n$, so we can estimate it using nearest neighbours in the same way that we
do for mean and variance.

To evaluate the performance of the model for different values of the smoothing parameter
we need the likelihood.

The density is just the multivariate normal:
\begin{equation}
 f=\frac{1}{(2\pi)^{\frac{p}{2}} D^\frac{1}{2}} \mbox{exp}\left(-\frac{1}{2}(z-\mu)^T\Sigma^{-1}(z-\mu)\right)
\end{equation}

where $z$ is the vector of $m$ anomalies along the track.

If we set $\mu=0$ this becomes

\begin{equation}
 f=\frac{1}{(2\pi)^{\frac{p}{2}} D^\frac{1}{2}} \mbox{exp}\left(-\frac{1}{2}z^T\Sigma^{-1}z\right)
\end{equation}

and the log likelihood over many independent tracks is:
\begin{equation}
\label{ll}
 \mbox{ln}f=\sum -\mbox{ln}[(2\pi)^{\frac{p}{2}} D^\frac{1}{2}]
           + \sum -\frac{1}{2}z^T\Sigma^{-1}z
\end{equation}

What is $\Sigma$?

Since the anomalies are normalised, the diagonal of $\Sigma$ is all 1's.

The first off diagonal is then given by the various values of $\phi_n$ along the track,
the second off diagonal by products of pairs of these $\phi$'s, and so on.

This gives:

\begin{center}
$
\left(%
\begin{array}{ccccc}
  1 & \phi_1 & \phi_1\phi_2 & \phi_1\phi_2\phi_3 & ... \\
  . & 1      & \phi_2       & \phi_2\phi_3       & ... \\
  . & .      & 1            & \phi_3             & ... \\
  . & .      & .            & 1                  & ... \\
  . & .      & .            & .                  & ... \\
\end{array}%
\right)
$
\end{center}

In the case where we assume that $\phi$ is constant everywhere, this then simplifies to

\begin{center}
$
\left(%
\begin{array}{ccccc}
  1 & \phi   & \phi^2 & \phi^3 & ... \\
  . & 1      & \phi   & \phi^2 & ... \\
  . & .      & 1       & \phi  & ... \\
  . & .      & .       & 1     & ... \\
  . & .      & .       & .     & ... \\
\end{array}%
\right)
$
\end{center}

So to calculate the likelihood we have to:
\begin{itemize}
    \item loop over the observed tracks
    \item for each track...
    \item find all the $\phi$'s on that track using the nearest neighbour model for $\phi$
    \item calculate $\Sigma$
    \item calculate $\Sigma^{-1}$ (numerically)
    \item calculate $D$ (numerically)
    \item calculate $\mbox{ln}f$ for that track
    \item move on to the next track
    \item add up the ln$f$'s for all the tracks
\end{itemize}

\subsection{Simulation for AR(1)}

To simulate, we have to do the following:
\begin{itemize}
    \item generate the 1st point on the track from the genesis model (or fix it from observations)
    \item generate the 2nd point by using the mean track and the variance model
    \item generate the 3rd and subsequent points using the mean track, the variance model and the memory model
\end{itemize}

So for the 3rd point onward the simulation works as follows:
\begin{itemize}
    \item Find the mean track from the mean track model, save that
    \item Find the variances from the variance model, save that
    \item Find the $\phi_u$ and $\phi_v$ from the autocorrelation model
    \item Derive the $s_u$ and $s_v$ from the $\phi$'s using $s^2=(1-\phi^2)$
    \item Apply $u_{n+1}=\phi_u u_{n} + s_u \epsilon$
    \item Apply $v_{n+1}=\phi_v v_{n} + s_v \epsilon$
    \item Multiply $u_{n+1}$ and $v_{n+1}$ by the appropriate variances from the variance model
    \item Add on the appropriate mean from the mean track model
    \item And we have a new point on the track
\end{itemize}

Note that if we did simplify the model so that $\phi$ is constant in space then this would be much easier
since we could just simulate the whole track in one go.
But in the general model we don't know what the $\phi$ along the
track will be in advance, so we don't know the correlation matrix in advance.

\section{Legal statement}

SJ was employed by RMS at the time that this article was written.

However, neither the research behind this article nor the writing
of this article were in the course of his employment, (where 'in
the course of their employment' is within the meaning of the
Copyright, Designs and Patents Act 1988, Section 11), nor were
they in the course of his normal duties, or in the course of
duties falling outside his normal duties but specifically assigned
to him (where 'in the course of his normal duties' and 'in the
course of duties falling outside his normal duties' are within the
meanings of the Patents Act 1977, Section 39). Furthermore the
article does not contain any proprietary information or trade
secrets of RMS. As a result, the authors are the owner of all the
intellectual property rights (including, but not limited to,
copyright, moral rights, design rights and rights to inventions)
associated with and arising from this article. The authors reserve
all these rights. No-one may reproduce, store or transmit, in any
form or by any means, any part of this article without the
authors' prior written permission. The moral rights of the authors
have been asserted.

The contents of this article reflect the authors' personal
opinions at the point in time at which this article was submitted
for publication. However, by the very nature of ongoing research,
they do not necessarily reflect the authors' current opinions. In
addition, they do not necessarily reflect the opinions of the
authors' employers.

\bibliography{jewson}

\newpage

\newpage

\newpage

\newpage

\newpage

\newpage

\newpage

\newpage

\newpage
\newpage


\newpage

\newpage

\newpage

\newpage

\newpage

\newpage

\newpage

\newpage

\newpage

\newpage
\begin{figure}[!htb]
  \begin{center}
    \scalebox{0.8}{\includegraphics{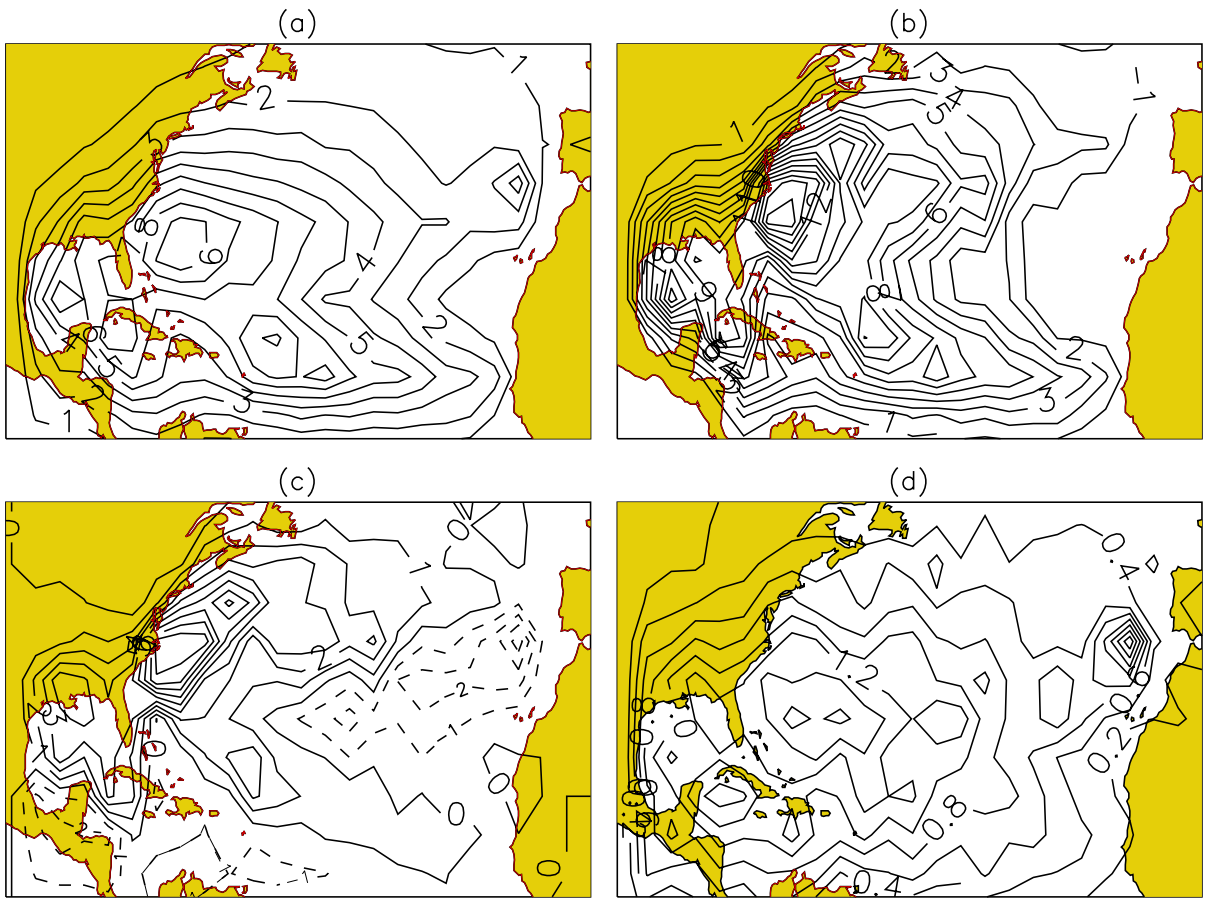}}
  \end{center}
  \caption{
  Track densities for model and observations.
  Panel (a) shows the track density for the AR(1) model, averaged over 34 realisations
  of 524 storms. Panel (b) shows the track density for observations, for 524 storms.
  Panel (c) shows the difference of these densities, and panel (d) shows the
  standard deviation of the density from the model (across the 34 simulations of 524 storms).
     }
  \label{density}
\end{figure}

\end{document}